\documentclass[aps,prd,preprint,groupedaddress]{revtex4-1}

\usepackage{graphicx}
\usepackage[table,xcdraw]{xcolor}
\usepackage{xcolor}
\usepackage{amsmath}
\begin{document}

\title{Quasinormal modes for Heavy Vector Mesons in a Finite Density Plasma}

\author{Nelson R. F. Braga}\email{braga@if.ufrj.br}
\author{Rodrigo da Mata}\email{rsilva@if.ufrj.br }
\affiliation{Instituto de F\'{\i}sica,
Universidade Federal do Rio de Janeiro, Caixa Postal 68528, RJ
21941-972 -- Brazil}


\begin{abstract} 
There are strong indications that ultra-relativistic heavy ion collisions, produced in accelarators, lead to the formation of a new state of matter: the quark gluon plasma (QGP).    
 This deconfined QCD matter is expected to exist  just for very short times after the collision. All the information one can get about the plasma is obtained from the particles that reach the detectors.  Among them, heavy vector mesons are particularly important. The abundance of $ c {\bar c}$ and $b {\bar b}$ states produced in a  heavy ion collision is a source of information about the plasma. In contrast to the light mesons, that completely dissociate when the plasma is formed,  heavy mesons  presumably  undergo a partial thermal dissociation.  The dissociation degree depends on the temperature and  also on the presence of magnetic fields and on the density (chemical potential). So, in order to get  information about the plasma out of the  quarkonium abundance data, one needs to resort to models that  provide the dependence of the  dissociation degree on these factors. 
Holographic phenomenological models provide a nice description for charmonium and bottomonium quasi-states in a plasma. In particular, quasi-normal modes associated with quarkonia states have been studied recently for a plasma with magnetic fields. Here we  extend this analysis of quasinormal models to the case when charmonium and bottomonium  are inside a  plasma with finite density. The complex frequencies obtained are then compared with a Breit Wigner approximation for the peaks of the corresponding thermal spectral functions, in order to investigate the quantitative agreement of the different descriptions of quarkonium quasi-states. 

\end{abstract}

\keywords{}

\maketitle

\section{ Introduction }    
One of the most fascinating challenges currently faced by physicists  is to build a detailed picture of the quark gluon plasma (QGP) from the particles that reach the detectors in heavy ion collisions. Hadronic matter, at the very high energy densities produced in these processes,  apparently produce the QGP, a state of matter where color is not confined inside bound states. It is not possible to observe the plasma directly. All the available information comes from particles created after this very short-lived state of matter  hadronizes and disappears. For reviews about QGP, see for example \cite{Bass:1998vz,Scherer:1999qq,Shuryak:2008eq,CasalderreySolana:2011us}. 

An important source of indirect information about the QGP is the abundance of heavy vector mesons. In contrast to light mesons, that dissociate once the plasma is formed, mesons made of $ c {\bar c}$ or  $b {\bar b}$ quarks may survive in the thermal medium. They undergo a partial dissociation process that depends on the flavor (charm or bottom), on the temperature and density of the plasma and on the presence of magnetic fields produced by the motion of charged particles. It is important to understand the thermal behavior of heavy vector mesons in a plasma in order to relate their relative abundance with the  properties of the medium. 

In recent years, it has been shown that holographic phenomenological models are a fruitful  framework for describing heavy vector mesons inside a thermal medium. In ref. \cite{Braga:2015jca} a holographic AdS/QCD model was proposed in order to represent the spectra of masses and decay constants of charmonium and bottomonium states in the vacuum. This model has two energy parameters, fixed by the criteria of the best fit to the experimental data. The extension to finite temperature and density appeared    in refs.  \cite{Braga:2016wkm,Braga:2017oqw} where the  spectral functions representing the quasi-states of heavy vector mesons were calculated.

Then, it was shown that a better fit for the spectra is obtained from an improved model involving three energy parameters \cite{Braga:2017bml,Braga:2018zlu}. These parameters represent:  the quark mass,  the string tension  and an  ultraviolet (UV) energy scale.  This UV energy parameter is related to the large mass change that happens in  non-hadronic decays and is necessary in order  to fit the  decay constant spectra.  A very nice picture of the dissociation process of charmonium and bottomnium states was obtained in these references by calculating the thermal spectral functions. 
 
A complementary tool for studying the dissociation process using holographic models is the determination of the quasinormal modes. The spectral function is obtained from real frequency solutions of the equations of motion of gravity fields dual to the hadrons. In contrast, quasinormal modes are solutions with complex frequency and more restrictive boundary conditions. The real part of the frequency is related to the thermal mass and the imaginary part to the thermal width of the quasi-states. 
In ref. \cite{Braga:2018hjt}, quasinormal modes for heavy vector mesons were studied using the holographic model of 
 \cite{Braga:2017bml,Braga:2018zlu}.  This analysis was then extended, for the case when a magnetic field is present, in ref. \cite{Braga:2019yeh}. The dissociation process corresponds to an increase in the imaginary part of the frequency of the quasinormal modes.  
 
 Here we present an extension of the analysis of heavy vector meson quasinormal modes to the case when the medium has a finite density. We investigate the dependence on the density of the complex frequencies for bottomonium and charmonium quasi-states at different temperatures. Then we compare the results obtained for the quasinormal frequencies  with the results form the spectral function approach. The comparison is performed by considering a Breit Wigner approximation for the spectral function peaks. 
 Such an approximation provides estimates for the real and imaginary parts of the quasinormal frequencies that are in a very nice agreement with the results obtained directly. This shows that the two approaches are consistent not only qualitatively but also from a quantitative point of view. 
 
 This letter is organized as follows: in section II  we review the holographic model for heavy vector mesons. In section III   we  present the calculation of quasinormal modes at finite temperature and density. Then in section IV  we compare the results obtained with a Breit Wigner approximation of the first spectral funcion peaks. Section  V is devoted to conclusions and final comments.

\section{Holographic model for heavy vector mesons}

In the phenomenological model developed in Refs.  \cite{Braga:2017bml,Braga:2018zlu}, vector mesons are described
by a vector field $V_m = (V_\mu, V_z) (\mu = 0, 1, 2, 3)$, representing the gauge theory current $J_\mu =\bar{\psi}\sigma^\mu\psi$ in the gravity side of the duality. The fields live in a curved five dimensional space, that is anti-de Sitter space for the case when the mesons are in the vacuum. Additionally, there is a scalar background. The action for the vector field reads
\begin{equation}
I \,=\, \int d^4x dz \, \sqrt{-g} \,\, e^{- \phi (z)  } \, \left\{  - \frac{1}{4 g_5^2} F_{mn} F^{mn}
\,  \right\} \,\,, 
\label{vectorfieldaction}
\end{equation}
where $F_{mn} = \partial_m V_n - \partial_n V_m$. The model contains three energy parameters that are introduced through the background scalar field $\phi(z)$:  
\begin{equation}
\phi(z)=\kappa^2z^2+Mz+\tanh\left(\frac{1}{Mz}-\frac{\kappa}{ \sqrt{\sigma}}\right)\,.
\label{dilatonModi}
\end{equation}

The parameter  $\sigma$,  with dimension of energy squared,  represents effectively the string tension of the strong quark anti-quark interaction.
 The mass of the heavy quarks is represented by $ \kappa$.  
  The third parameter, $M$,  has a more subtle interpretation.
Heavy vector mesons undergo non hadronic decay processes, when the final state consists of light leptons, like an $e^+ e^-$ pair. In such transitions  there is a  very large mass  change, of the order of the meson mass.  The parameter $ M$  represents effectively the mass scale of such a transition, characterized by a matrix element  $ \langle 0 \vert \, J_\mu (0)  \,  \vert n \rangle = \epsilon_\mu f_n m_n \, $, where $f_n $ is the decay constant, $ \vert n \rangle $ is a meson state  at radial excitation level $n$ with mass $m_n$ , $  \vert 0  \rangle $ is the hadronic vacuum and    $J_\mu$ the hadronic current.  The values that provide the best fit to charmonium and bottomonium spectra of masses and decay constants at zero temperature are respectively
\begin{eqnarray}
\kappa_c&  = 1.2, \sqrt{\sigma_c } = 0.55, M_c=2.2 \, ; \, 
\kappa_b& = 2.45, \sqrt{\sigma_b } = 1.55, M_b=6.2  \,,
  \label{parameters}
  \end{eqnarray}     
\noindent where all quantities are expressed in GeV. 
The geometry that corresponds to  a thermal medium with a finite density $\mu$ was studied in refs.\cite{Colangelo:2010pe,Colangelo:2011sr,Colangelo:2012jy}. It is a 5-d anti-de Sitter charged black hole space-time  with the metric 
 \begin{equation}
   ds^2 \,\,= \,\, \frac{R^2}{z^2}  \,  \Big(  -  f(z) dt^2 + d\vec{x}\cdot d\vec{x}  + \frac{dz^2}{f(z) }    \Big)   \,,
 \label{metric2}
\end{equation}
where  
\begin{equation}
 f (z) = 1 - \frac{z^4}{z_h^4} - q^2 z_h^2 z^4+q^2 z^6 ,
\end{equation}
 and $z_h$ is the horizon position where $f(z_h) = 0$. The relation between $z_h$ and the temperature $T$ of the black hole, assumed to be the same as the gauge theory temperature,  comes from the condition of absence of conical singularity at the horizon:
\begin{equation} 
T =  \frac{\vert  f'(z)\vert_{(z=z_h)}}{4 \pi  } = \frac{1}{\pi z_h}-\frac{  q^2z_h ^5}{2 \pi }\,.
\label{temp}
\end{equation}

The parameter $q$ is proportional to the black hole charge. In the dual gauge theory, $q$ is related to the density of the medium, or  quark chemical  potential, $\mu$. In the gauge theory Lagrangean, 
  $\mu$ would appear multiplying the quark density $  \bar{\psi}\gamma^0 \psi \,$.  
So it  would work as the source of correlators for   this operator. In the dual supergravity description the time component $V_0$ of the vector field plays this role. 
So, one considers a particular solution for the vector field   $ V_m $ with only one 
non-vanishing component: $  V_0 = A_0 (z) $  ($V_z =0, V_i = 0$). Assuming that the relation between $q $ and $\mu$ is the same as in the case of no background, that means $ \phi (z)  = 0$, the solution for the time component of the vector field is:
 $ A_{0}(z)=c - qz^2 $, where $c$ is a a constant. Imposing  $A_{0}(0) = \mu $ and $A_{0}(z_h)=0$ one finds: 
\begin{equation}
\mu = q z_h^2  \,.
\label{chemicalpotential}
\end{equation}  
From eqs. (\ref{temp}) and (\ref{chemicalpotential}) it becomes clear that specifying both $z_h$ and $q$, the values of the temperature and the chemical potential are fixed and contained into the metric  (\ref{metric2}).


\section{ Quasinormal modes at finite density}

In the context of gauge/gravity duality, quasinormal modes are normalizable gravity solutions representing the quasi-particle states in a thermal medium, with complex frequencies $\omega $. The real part, $\operatorname{Re}(\omega )$, is related to the thermal mass and the imaginary part, $\operatorname{Im}(\omega )$, is related to the thermal width. In the zero temperature limit one recovers the vacuum hadronic states, corresponding to solutions that are called normal modes. The normalizability condition requires that, either in the zero temperature case or in the finite temperature one, the fields must vanish at the boundary z=0. The main difference in that at finite $T$ there is an event horizon  where one additionally have to impose infalling boundary conditions. These two types of boundary conditions are simultaneously satisfied  in general  for complex frequencies.  


 In the radial gauge $V_z = 0,$ and considering solutions of the type $V_{\mu} = e^{-i\omega t+i k x_{3}}V_{\mu}(z,\omega,k)$ that corresponds to plane waves propagating in the  $x_3$ direction with wave vector $p_\mu = (-\omega,0,0,k)$. In terms of the electric field components $E_1 = \omega V_1, E_2 = \omega V_2$ and $E_3 = \omega V_3 + k V_t $  the equations of motion coming from action (\ref{vectorfieldaction}) with the metric(\ref{metric2}) take the form:

\begin{equation}
\label{eqzz}
E_{\alpha}''+\left(\frac{f'}{f}-\frac{1}{z}-\phi'\right)E_{\alpha}'+\frac{1}{f^2}\left(\omega^2 - k^2f \right)E_{\alpha}=0 \, , \,\,\,\,\,\,\,\,\,\,\ (\alpha=1,2)
\end{equation} 
  
\begin{equation}\label{eqx3x3}
E_{3}''+\left(\frac{\omega}{\omega^2-k^2f}\frac{f'}{f}-\frac{1}{z}-\phi'\right)E_{3}'+\frac{1}{f^2}\left(\omega^2 - k^2f \right)E_3=0, \,\,\,\,
\end{equation} 
where (') represents derivatives with respect to the radial z coordinate.

One has to impose the normalizability condition at $z=0$ and the infalling condition at $z=z_h$. 
The idea, in order to impose the boundary conditions at the horizon, is to re-write the field equations in such a way that they separate into a combination of  infalling  and outgoing waves. It is convenient to introduce the Regge-Wheeler tortoise coordinate $r_{*}$. This coordinate is defined by the
relation $\partial_{r_{*}} = - f(z)\partial_{z}$ with z in the interval $0 \leq z \leq zh$ and can be expressed explicitly by integrating the former relation and  imposing $r_{*}(0)=0$. In order to have a Schr\"{o}dinger like equation, one can perform a Bogoliubov transformation on the electric fields, introducing  $(\psi_j = e^{-\frac{B_j (z)}{2}}E_j)$. Then, Eqs. (\ref{eqzz}) and (\ref{eqx3x3}) reduce to the form:
\begin{equation}
\label{Sequation}
\partial^{2}_{r_*}\psi_j +\omega^{2}\psi_j = U_j\psi_j \,,
\end{equation}
\noindent with 
\begin{equation}
\label{meq}
B_{\alpha}(z) = \log(z) + \phi; B_3(z) = \log\left(\frac{w^2-k^2f}{\omega^2}z\right) +\phi .
\end{equation}

For $j = (1,2,3)$ these potentials diverge at $z=0$ so we must have $\psi_j(z=0)=0 $. At the horizon  we have for both potentials that $U_j(z=z_h)=0$  in the limit $z \to z_h$ one expects to find  \textit{infalling} $\psi_j = e^{-i\omega r_*}$ and \textit{outgoing} $\psi_j = e^{+i\omega r_*}$ wave solutions for equation (\ref{Sequation}). Only the first kind of solutions are physically allowed.    The Sch\"{o}dinger like  equation can be expanded near the horizon leading to the following expansion the field solution:
\begin{equation}
\psi_j=e^{-i\omega r_*(z)} \left[1+a^{(1)}_j\left(z-z_h\right)+a^{(2)}_j\left(z-z_h\right)^2+\dots\right] \,.
\label{expansion}
\end{equation}

One can solve recursively for $a^{(n)}_j$. The first coefficients are:
\begin{equation}
a^{(1)}_\alpha=   \frac{\left(2 - q^2 z_h^6 \right) \left(z_h \left(\frac{k^2}{2-q^2
   z_h^6}+2 \kappa ^2\right)-\frac{\text{sech}^2\left(\frac{\kappa
   }{\sqrt{\sigma }}-\frac{1}{M z_h}\right)}{M
   z_h^2}+\frac{1}{z_h}+M\right)}{2 \left(q^2 z_h^6+i \omega 
   z_h-2\right)} \,,
\end{equation}

\begin{equation}
a^{(1)}_3= \frac{\left( 2 - q^2 z_h^6 \right)  
\left(z_h \left(\frac{k^2}{2-q^2
   z_h^6}+2 \kappa ^2\right)
    -   \frac{\text{sech}^2\left(\frac{\kappa
   }{\sqrt{\sigma }}-\frac{1}{M z_h}\right)}{M
   z_h^2} 
      +\frac{ 4 k^2 + \omega^2 - 2 k^2 q^2 z_h^6 }{\omega^2 z_h} + M \right)  } 
     {2 \left(q^2 z_h^6+i \omega 
   z_h-2\right)} \,.
\end{equation}

Eqs. (\ref{eqzz}) and (\ref{eqx3x3}) are solved numerically for complex frequencies using  a method that consists in imposing infalling boundary conditions for the electric field at the horizon. These  conditions are obtained by expressing the expansion (\ref{expansion})  in terms of the field $E$  near the horizon:  
\begin{equation}
\displaystyle \lim_{z\to\ z_h} E_j(z) \longrightarrow e^{-i\omega r_*(z) + \frac{B_j(z)}{2}} \left(1+a^{(1)}_j\left(z-z_h\right)+a^{(2)}_j\left(z-z_h\right)^2+\dots\right) \,.
\end{equation}
This expansion leads to the following boundary conditions for the field and it's derivative at the horizon: 
\begin{eqnarray}
E_j(z_h) &=&  e^{-i\omega r_*(z_h) + \frac{B_j(z)(z_h)}{2}} \,, \\
E_j^{'}(z_h)  &=&  \left(-i\omega r'_{*}(z_h) + \frac{B'_j(z_h)}{2}+a^{(1)}_j\right) E_j(z_h) \,.
\end{eqnarray} 
Then one searches for the complex frequencies that provide solutions vanishing on the boundary:   $E_j (z=0)=0$. These are the  quasinormal frequencies and the corresponding -- normalizable -- solutions  are the quasinormal modes, that represent the heavy meson quasi-states in the thermal medium.

\noindent {\bf Results } 

Let us start with heavy mesons at rest in the medium. Figures {\bf  1} and {\bf  2}  present,   for the first states of charmonium J/$\psi$ and bottomonium $\Upsilon$, respectively, the  dependence of the real and imaginary parts of the quasinormal frequencies on the temperature at four different values of the chemical potential $\mu$.

\begin{figure}[h!]
\label{qnmccbartemp}
\includegraphics[scale=0.36]{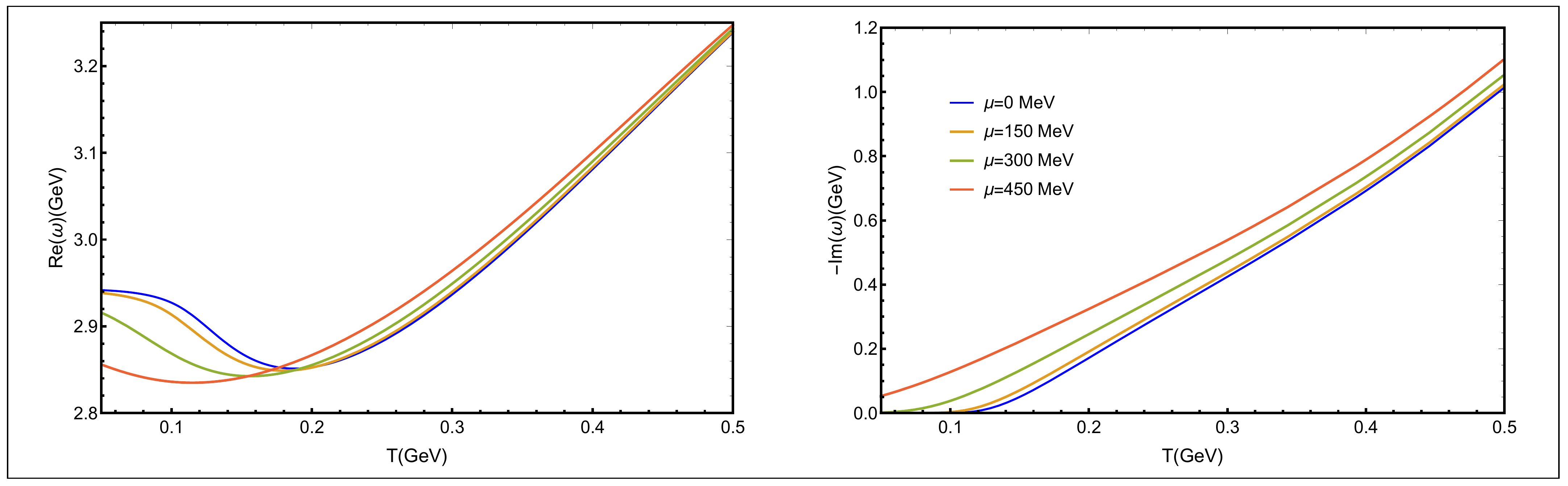}
\caption{Real  (left panel) and imaginary (rigth panel) parts of the  quasinormal mode frequency of  J/$\psi$ as a function of the temperature for four different values of the chemical potential: $\mu = 0, 150, 300, 450 $ MeV. }
\end{figure}

\begin{figure}[h!]
\label{qnmbbbartemp}
\includegraphics[scale=0.36]{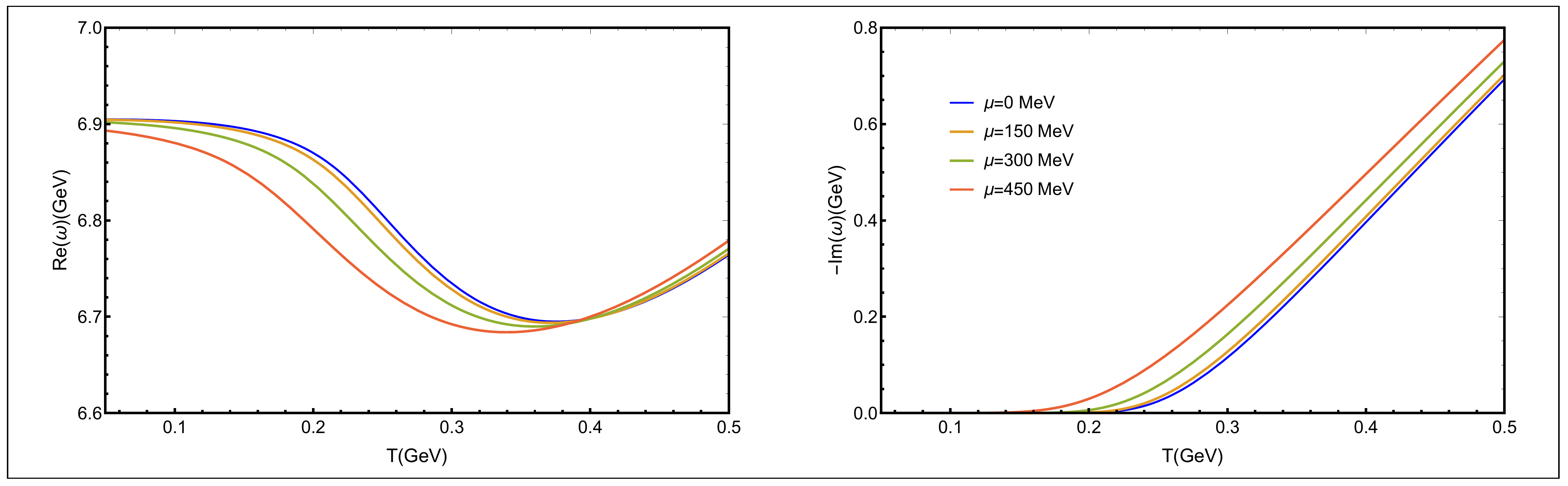}
\caption{ Real  (left panel) and imaginary (rigth panel) parts of the  quasinormal mode frequency of  $\Upsilon$  as a function of the temperature for four different values of the chemical potential: $\mu = 0, 150, 300, 450 $ MeV. }
\end{figure}

The behavior of charmonium and bottomonium is similar.  As one can see  on the left panels,  the thermal mass represented by    $\operatorname{Re}(\omega )$ decreases  with the chemical potential for low temperatures. For higher temperatures,  $T  \gtrsim   200 MeV$ for J/$\psi$ and $T \gtrsim   400 MeV$ for $\Upsilon$,  the effect of the chemical potential is the opposite but much smaller.  

The imaginary part of the frequency  $\operatorname{Im}(\omega )$  is   related to the width of the associated  peak of the spectral function. An increase in 
$\operatorname{Im}(\omega )$ corresponds to a  broadening of the peak. 
On the other hand, the broadening of a peak corresponds to  a smaller  probability of finding the quasiparticle in the medium because of the dissociation.  The decrease in this probability is what we call as an increase in the ``dissociation degree''.  For both charmonium and bottomonium ground states there is a monotonic increase in the absolute value of  $\operatorname{Im}(\omega )$  with  temperature and chemical potential. 
  This means that, the denser is the plasma, the higher is the dissociation degree of heavy vector mesons. 
  
 Now, let us consider mesons in motion in the plasma. 
 Figure { \bf 3}   show the dependence  of the charmonium  J/$\psi$  quasinormal frequencies at $ T $ = 125 MeV on the momentum $k$. The upper panels represent the case when the motion is transverse to the polarization and the lower panels show the case when the momentum is longitudinal to the polarization. Figure {\bf 4} shows the analogue results for the bottomnium $\Upsilon$ quasi-state at the temperature $ T$  = 300 MeV. The reason for choosing these temperatures is that the corresponding spectral functions, for   charmonium  at $ T $ = 125 MeV and for  bottomonium at  $ T$  = 300 MeV,  present clear peaks for the first quasi-state. So, at this temperatures it is simple to  observe the effect of the density.

 \begin{figure}[ht]
\label{dispTccbar}
\includegraphics[scale=0.37]{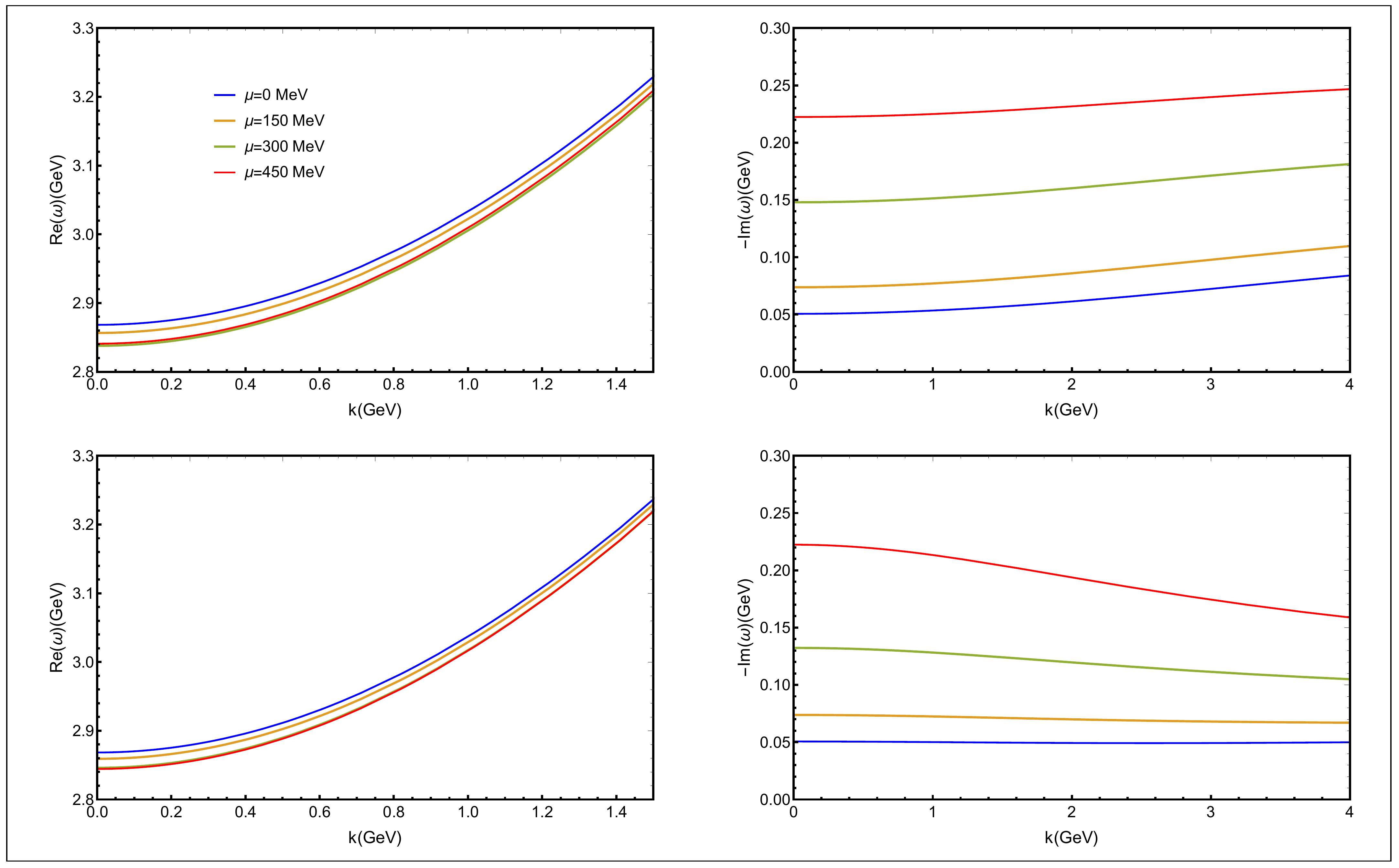}
\caption{Dispersion relations for J/$\psi$ QNM at $T $= 125 MeV. The upper panels show the transverse  ($j=1,2$) case and the bottom panels the longitudinal ($j=3$) one.  The real part of the frequencies are on the left panels and the imaginary part on the right panels.}

\end{figure}
\begin{figure}[ht]
\label{dispTbbar}
\includegraphics[scale=0.36]{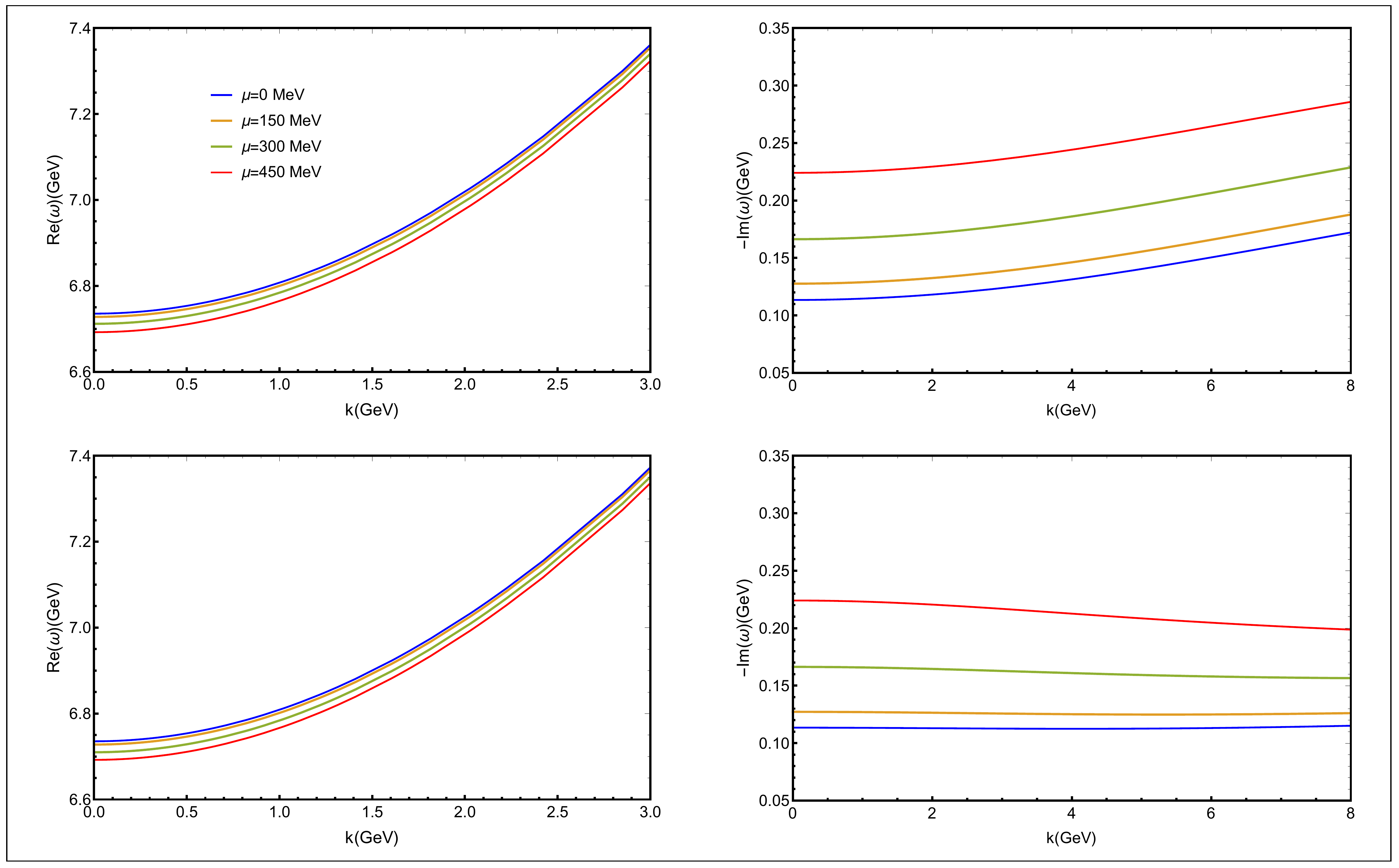}
\caption{Dispersion relations for $\Upsilon$ QNM at $T $ = 300 MeV. The upper panels show the transverse  ($j=1,2$) case and the bottom panels the longitudinal ($j=3$) one.  The real part of the frequencies are on the left panels and the imaginary part on the right panels.}
\end{figure}

Both flavors  present a similar  behavior. The real parts of the QNM modes  represent  for $k=0$ the thermal mass. The dependence on $ k$ is approximately equal to the trivial dispersion relation for a free  relativistic particle:  $ w = \sqrt{m^2 + k^2 }$. 
For the imaginary parts of the QNM, the  dispersion relation is shown on the panels on the right-hand side of figures {\bf 3} and {\bf 4}.  For both flavors the dissociation considerably increases with the chemical potential but the effect of the motion of the meson inside the plasma depends on the direction of the momentum with respect to the polarization. 
For motion transverse to polarization the dissociation degree increases with the momentum $k$ while for  longitudinal motion it decreases with the momentum. This effect is more 
noticeable for higher chemical potentials, when the dissociation caused by the density is partially compensated by longitudinal motion. 

\section{Comparison with spectral function results }
 
Quasinormal modes (QNM) and spectral functions are complementary  approaches for studying  the thermal behavior of heavy vector mesons in a plasma. It is interesting to check if the results obtained here using QNM are consistent with the spectral function description, using the same holographic approach,  presented in ref. \cite{Braga:2017bml}. It is known that an increase in the imaginary part of the quasinormal frequencies corresponds to a broadening  of the corresponding peak of the thermal spectral function. Such a qualitative analysis is trivial and indicates consistency between the two approaches. Much more interestingly, it is possible to make a quantitative test of consistency. 
 In the vicinity of the peaks, the spectral function calculated from the imaginary part of the current-current retarded propagator has the approximate Breit Wigner (BW) form:

\begin{equation}
    \frac{\rho(\omega)}{\omega}\approx \frac{a (\Gamma/2)^2 }{(\omega- M)^2+(\Gamma /2)^2}\,,
    \label{BWA}
\end{equation}
\noindent where the height of the peak $a$, the thermal mass $M$ and the width $\Gamma$  can be determined by the numerical adjust of  the spectral function $\rho(\omega)$  (see for example \cite{Ceci:2013zta} for a discussion).

Figure {\bf 5} shows the  spectral functions calculated from the model considered here  for charmonium at $T =125 MeV$ (left panel)  and bottomonium at  $T= 300 MeV$ (rigth panel). For details on how to determine  these spectral functions,  see ref. \cite{Braga:2017bml}. 

\begin{figure}[ht]
\label{spectralfunc}
\includegraphics[scale=0.36]{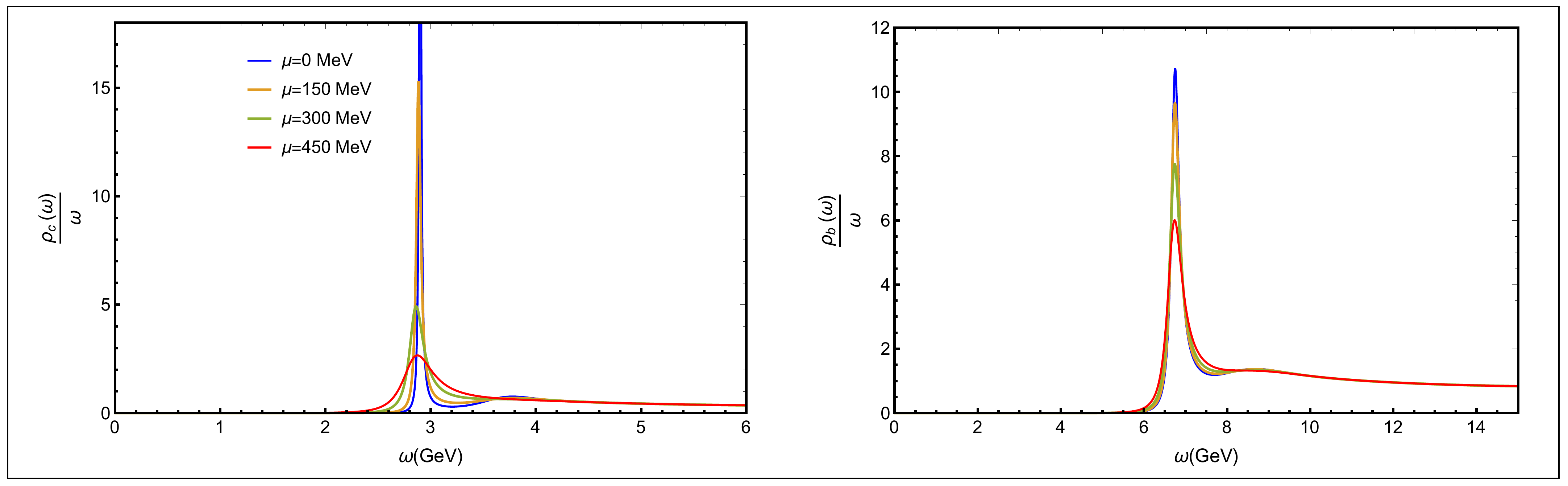}
\caption{Spectral functions for charmonium at $T $ = 125 MeV and bottomonium at $T $ = 300 MeV}
\end{figure}

\begin{figure}[ht]
\label{BWadjust}
\includegraphics[scale=0.36]{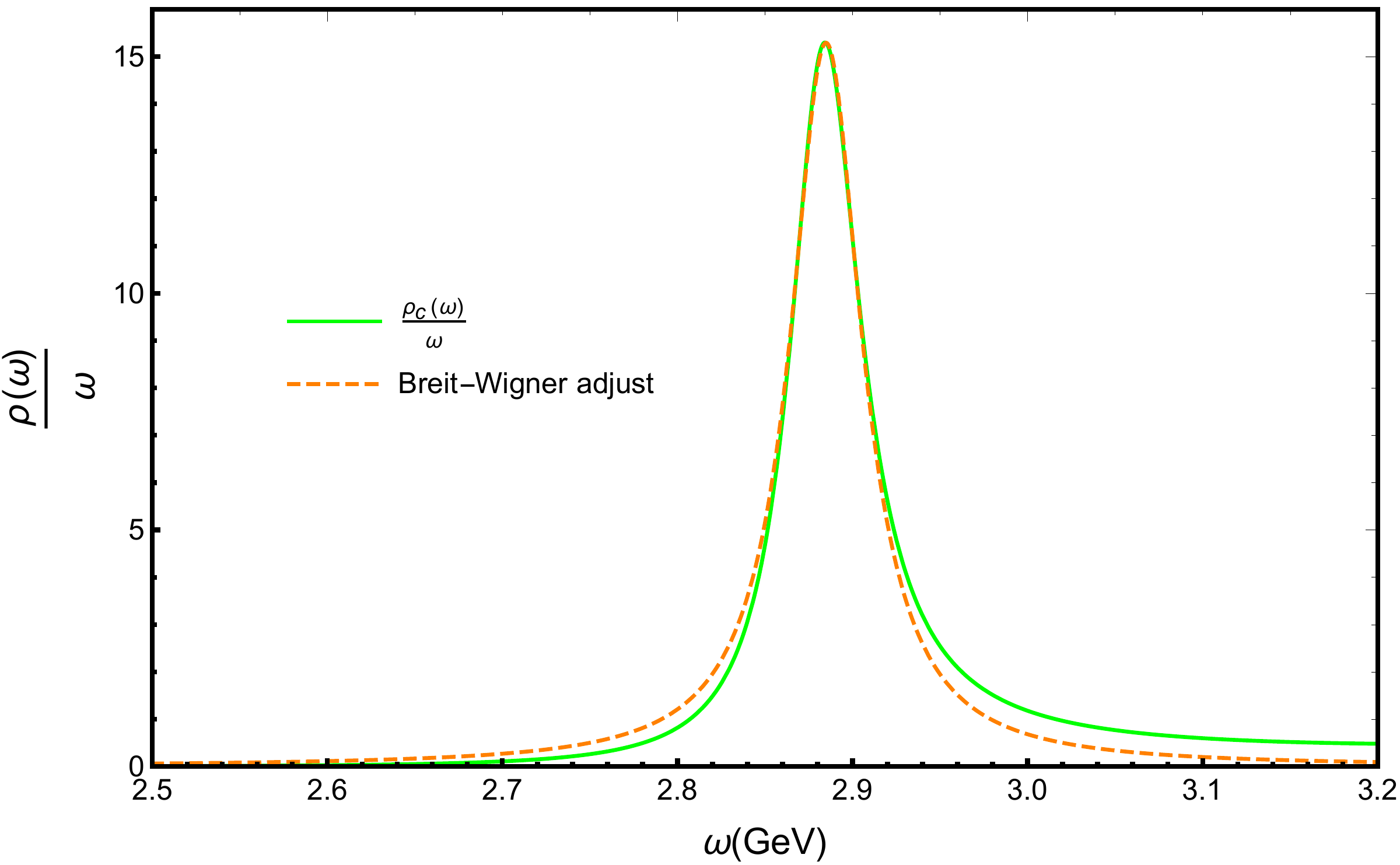}
\caption{ Breit-Wigner fit compared  with the actual spectral function peak  for charmonium at $T $ = 125 MeV and $\mu $ = 150 MeV  }
\end{figure}

It is clear from these plots that the heigth of the peaks decrease with the chemical potential.  In order to illustrate how the BW approximation works, we  show in figure { \bf 6}  the BW  fit and the actual peak for charmonium at $T $ = 125 MeV and $\mu = $ 150 MeV 
One notices that near the peak the BW approximation works very well.  
As mentioned  in the previous section, the reason for choosing these  temperatures is that the corresponding spectral functions, for   charmonium  at $ T $ = 125 MeV and for  bottomonium at  $ T$  = 300 MeV, present very clear peaks for the first quasi-states. So, at this temperatures it is simple to  observe the effect of the density. 

Performing a numerical fit of spectral functions, near the first peak,  to the Breit Wigner (BW)  form (\ref{BWA}) one can test the agreement with the quasinormal modes approach.  The comparison  comes from the identification of the thermal mass $M$ with the real part of the QNM frequency  $\operatorname{Re}(\omega )$ and the resonance width $\Gamma $ with twice the absolute value of the  imaginary part: $2  \vert \operatorname{Im}(\omega ) \vert $. 

Tables I and II summarize the results for QNM frequencies calculated  using the method presented at section III  and the results for the corresponding quantities obtained from a  Breit-Wigner fit of the spectral functions peaks. 
One notices that the relative discrepancies are very small and then concludes that the results of quasinormal modes and spectral functions are   consistent. They represent complementary approaches to study the thermal behavior
of quarkonium inside a plasma.

It is important to remark that similar results are obtained if one chooses different temperatures. The choice of $ T $ = 125 MeV for  charmonium  and  $ T$  = 300 MeV for  bottomonium was motivated by the fact that at these temperatures 
the effect of the density is more evident. However the same kind of agreement between the spectral function and quasinormal mode approaches is observed at different temperatures.

\begin{table}[ht]
\begin{tabular}{|l|l|l|l|l|l|}
\hline
Quasi-state & $ \mu $ (MeV)  & $ T $ (MeV)  & $ \operatorname{Re}(\omega )  $ (MeV)   & $ M $ (MeV ) & rel. discr.  (\%)\\ \hline
J/$\psi$ &  0  &  125    & 2.89931    & 2.88558  &1.8 \\ \hline
J/$\psi$  &  150  &  125   & 2.88270  & 2.86565  & 0.54  \\ \hline
J/$\psi$  &  300  &  125   & 2.85034  & 2.88491  & 0.076 \\ \hline
J/$\psi$   & 450   &  125    & 2.83546  & 2.89992  & 0.021  \\ \hline
\hline
$\Upsilon$  &  0  &  300  & 6.73452  & 6.75621 & 0.32 \\ \hline
$\Upsilon$ &  150  &  300  & 6.72812  & 6.75463   & 0.39 \\ \hline
$\Upsilon$  &  300   &  300  & 6.71170 & 6.74551  & 0.50  \\ \hline
$\Upsilon$ &  450   &  300  & 6.69218  & 6.74412 & 0.78 \\ \hline
\end{tabular}
\caption{ Real part of the quasinormal frequency  $\operatorname{Re} (\omega )$, compared with the thermal mass $M$ obtained from  Breit-Wigner fit of the spectral function  and their relative discrepancy.   }
\end{table}

\begin{table}[ht]
\begin{tabular}{|l|l|l|l|l|l|}
\hline
Quasi-state & $ \mu $ (MeV)  & $ T $ (MeV)  &  $ \vert  \operatorname{Im}(\omega )  \vert $ (MeV)   & $   \Gamma /2 $ (MeV ) & rel. discr.  (\%)\\ \hline
J/$\psi$ &  0  &  125     & 0.0108540    & 0.0108923  &0.35 \\ \hline
J/$\psi$  &  150  &  125   & 0.0246513  & 0.0248886  & 0.96    \\ \hline
J/$\psi$  &  300  &  125  & 0.0805289  & 0.0804919   & 0.046    \\ \hline
J/$\psi$   & 450   &  125   & 0.173714  & 0.173152  & 0.32   \\ \hline
\hline
$\Upsilon$  &  0  &  300    & 0.115284   & 0.115169 & 0.10   \\ \hline
$\Upsilon$ &  150  &  300  &0.127202  & 0.13211  & 3.8  \\ \hline
$\Upsilon$  &  300   &  300   &0.163493  & 0.164364  & 0.53   \\ \hline
$\Upsilon$ &  450   &  300   &0.224074 & 0.229888   & 2.6   \\ \hline
\end{tabular}
\caption{ Absolute value of the imaginary  part of the quasinormal frequency   $ \vert \operatorname{Im}(\omega ) \vert $, compared with one half of the thermal width  $\Gamma $ obtained from  Breit-Wigner fit of the spectral function and their relative discrepancy. }
\end{table}

\section{Conclusions} 
 In this letter the behavior of heavy vector mesons inside a plasma with finite chemical potential was studied through  the determination of the quasinormal modes. A holographic model was used,  in order  to describe the quasi-states of  $c {\bar c} $ and $b {\bar b} $  
 in terms of normalizable field solutions in a five dimensional dual space.  This dual geometry contains a black hole with a Hawking temperature assumed to be the same as the gauge theory temperature. The chemical potential, or density, of the medium is represented in the holographic description, by the charge of the black hole. 
 
 The results obtained show how the density affects the partial thermal dissociation and the thermal mass of the quarkonia in the medium. In particular, the higher the density, the higher the dissociation degree. This fact holds for heavy mesons at rest or in motion relative to the medium. However, there is a non trivial aspect. For motion in the direction perpendicular to the polarization (transverse motion) the dissociation degree increases with momentum. In contrast, for motion in the direction of polarization  (longitudinal motion), the dissociation degree decreases with momentum. So, for this longitudinal case, motion and density have opposite effects. 
 
 Regarding the thermal mass, there is also a non trivial behavior. For low temperatures 
 ($T  \lesssim   200 MeV$ for J/$\psi$ and $T \lesssim   400 MeV$ for $\Upsilon$)
 the thermal mass decreases with the chemical potential. For higher temperatures, the mass increases slightly with the chemical potential.  
 It is important to note that the plasma at $T$ = 125 MeV considered here would be a super-cooled plasma phase if the critical deconfinement temperature is above this value. 
 
 There are many interesting previous studies of mesons using holography to describe thermal effects and heavy flavors like for example  \cite{Kim:2007rt,Fujita:2009wc,Fujita:2009ca,Branz:2010ub,Gutsche:2012ez,Fadafan:2011gm,Fadafan:2012qy,Fadafan:2013coa,Afonin:2013npa,Mamani:2013ssa,Hashimoto:2014jua,Liu:2016iqo,Liu:2016urz,Ballon-Bayona:2017bwk,Braga:2017fsb,Dudal:2018rki,Braga:2018fyc,Gutsche:2019blp,Bohra:2019ebj,Zhang:2019qhm}.     
An alternative approach to describe heavy meson spectra was proposed very recently in ref. \cite{MartinContreras:2019kah}.

\noindent {\bf Acknowledgments:}  The authors are supported by  CNPq - Conselho Nacional de Desenvolvimento Cientifico e Tecnologico (N.B by grant  307641/2015-5 and R. M. by a graduate fellowship). This work received also support from  Coordenação de Aperfeiçoamento de Pessoal de Nível Superior - Brasil (CAPES) - Finance Code 001.

 \end{document}